\documentclass[letterpaper,peerreview,draftcls,onecolumn,11pt]{IEEEtran}
\usepackage{amsmath, amssymb, amsthm, graphicx, marvosym, cite, psfrag}

\def\ts{\textstyle}
\def\binomial{\mathop{\mathrm{binomial}}}

\def\Half{{\frac{1}{2}}}

\def\var{\mathop{\mathrm{var}}}
\def\E{\mathop{\mathrm{E}}}

\title{Benefiting from Disorder: \\ Source Coding for Unordered Data}
\author{Lav~R.~Varshney,~\IEEEmembership{Graduate Student Member,~IEEE,}
        and~Vivek~K~Goyal,~\IEEEmembership{Senior~Member,~IEEE}%
\thanks{This work was supported in part by an NSF Graduate Research Fellowship,
	NSF Grant CCR-0325774, 
        the Texas Instruments Leadership University Consortium Program, and
        the Centre Bernoulli at 
        \'{E}cole Polytechnique F\'{e}d\'{e}rale de Lausanne.}%
\thanks{The material in this paper was presented in part at
        the Information Theory and its Applications Inaugural Workshop,
        La Jolla, California, February 2006;
        the IEEE Data Compression Conference, Snowbird, Utah, March 2006;
        and the 2007 Information Theory and its Applications Workshop,
        La Jolla, California, January/February 2007.}%
\thanks{L.~R. Varshney (email: lrv@mit.edu) and
        V.~K Goyal (email: vgoyal@mit.edu) are with
        the Department of Electrical Engineering and Computer Science and
        the Research Laboratory of Electronics,
        Massachusetts Institute of Technology.
        L.~R. Varshney is also with the Laboratory for Information and Decision
        Systems, Massachusetts Institute of Technology.}}

\newtheorem{thm}{Theorem}
\newtheorem{lemma}{Lemma}

\begin{document}

\maketitle

\begin{abstract}
The order of letters is not always relevant in a communication task.
This paper discusses the implications of order irrelevance on
source coding, presenting results in several major branches of source
coding theory:
lossless coding, universal lossless coding, rate-distortion,
high-rate quantization, and universal lossy coding. 
The main conclusions demonstrate that there is a significant
rate savings when order is irrelevant.  In particular, lossless coding of
$n$ letters from a finite alphabet requires $\Theta(\log n)$ bits and
universal lossless coding requires $n + o(n)$ bits for many countable alphabet sources.
However, 
there are no universal schemes that can drive a strong redundancy measure 
to zero.
Results for lossy coding include distribution-free expressions for the
rate savings from order irrelevance in various high-rate quantization schemes.
Rate-distortion bounds are given, and it is shown that the analogue of
the Shannon lower bound is loose at all finite rates.
\end{abstract}

\begin{IEEEkeywords}
compression, lossless coding, multisets, order statistics,
quantization, rate-distortion theory, universal coding, types
\end{IEEEkeywords}

\newpage

``ceiiinosssttuv'' 
\hspace*{\fill}
\emph{--- Robert Hooke in 1676, establishing priority for the statement} \\
\hspace*{\fill}
\emph{``ut tensio sic vis''
      (as is the extension, so is the force)
      published in 1678~\cite{Moyer1977}.}

\section{Introduction}
Are there situations where {\tt claude shannon} is no different than
{\tt a sound channel}; where {\tt maximum entropy} might be
considered a reasonable reconstruction for {\tt momentary mixup}?\footnote{Anagrams due to R.~J. McEliece, 2004 Shannon Lecture.}  
If one is interested in communicating a textual source, an anagram is not a sufficient 
representation.  If however, one is simply interested in classifying the language, 
an anagram may be sufficient because it gives the same first-order language 
approximation.\footnote{In Shannon's sense of language approximation~\cite{Shannon1948}, 
first-order approximation requires that the distribution of letters matches the source, 
second-order approximation requires that the distribution of digrams matches the source,
and so on; 
see also classical criticisms to this method of approximation~\cite{Chomsky1956}.} 
That is to say,
the values of the letters (a multiset) may be important 
when the order of the letters (a permutation) is not.
The order of source letters is irrelevant in
a multitude of scenarios beyond such 
language representation and related texture representation applications~\cite{EfrosL1999}.
Examples include warehouse inventories~\cite{Lempel1986}; 
records in scientific or financial databases~\cite{BuchsbaumFG2003}; 
collections of multimedia files; arrival processes~\cite{Rubin1974};
visual languages that have multiset grammars~\cite{MarriottM1998};
data in a parallel computing paradigm~\cite{BanatreM1993}; and
the channel state in chemical channels~\cite{PermuterCVW2006}.
Moreover, it has been suggested that when humans use data for 
recognition or recall tasks~\cite{JuddS1959},
or for judgments of coincidences~\cite{GriffithsT2001,GriffithsT2007},
the order of symbols is not relevant.

If a sample consists of independent observations from the same
distribution, then associated minimum variance unbiased estimators are 
symmetric in the observations~\cite{Rao1965}.  Therefore when coding 
for estimation, the multiset of observations 
is all that need be represented, cf.~\cite{HanA1998}.  
Moving beyond the point-to-point case, in distributed inference,
often particle-based~\cite{ArulampalamMGC2002,IhlerFMW2005} and 
kernel-based~\cite{Parzen1962,IhlerFW2004}
representations of densities must be communicated.  As
\[
p(x) = \sum_{i} \phi\left(x - x_i\right)
     = \sum_{i} \phi\left(x - x_{\pi(i)}\right)
\]
for any permutation $\pi(\cdot)$, the multiset of representation 
coefficients $\{x_i\}$ may be communicated rather than the 
sequence of these values $(x_i)$.  This extends to any 
destination that performs permutation-invariant computations.

The aim of this paper is to develop ramifications of order irrelevance
on source coding problems.  We consider lossless coding, universal lossless coding,
high-rate and low-rate quantization, and rate distortion theory.
In all of these, the rate requirement is obviously reduced by making order irrelevant.
The reduction can be dramatic:
for lossless coding of $n$ symbols,
the required rate is changed from $O(n)$ to $O(\log n)$;
for lossy coding, with large enough $n$ arbitrarily small mean squared error
(MSE)
can be achieved with zero rate.
These examples are made precise in the sequel.

\subsection{Notation and Formalism}
We consider the encoding of multisets and sequences of letters of size $n$
drawn from an alphabet $\mathcal{X}$.
When $\mathcal{X}$ is discrete, we take it to be
the (possibly-infinite) set $\{1,\,2,\,\ldots,|\mathcal{X}|\}$ without
loss of generality.
In addition to standard uses of parentheses, we will also use 
parentheses and braces to distinguish between sequences and multisets.
For the ordered sequence $X_1,\,X_2,\,\ldots,\,X_n$
that is often denoted $X_1^n$ in the
information theory literature, we write $(X_i)_{i=1}^n$.
When these $n$ symbols are taken as an unordered multiset,
we write $\{X_i\}_{i=1}^n$.  The range limits are often omitted.  We will
refer to the distribution that describes $(X_i)_{i=1}^n$
as the parent distribution. 

There are two perspectives that can be taken to relate
(order-irrelevant) source coding of $\{X_i\}_{i=1}^n$ with
(standard) source coding of $(X_i)_{i=1}^n$.
In either case we take the sample space $\Omega$ to be
the set of all sequences $\mathcal{X}^n$.
Under the first perspective, which we take when discussing lossless coding,
we define the event algebra, $\mathcal{F}$, based on 
permutation-invariant equivalence classes of sequences, rather than the 
sequences themselves.  Since events are defined in terms of multisets, 
the source coding problem is formally no different than the standard one,
though the results are interestingly different.  

For lossy coding, we take an alternative perspective where the event algebra
is based on sequences.  Order irrelevance is incorporated by considering
fidelity criteria with a permutation-invariance property. 
These fidelity criteria cannot be stated in single-letter terms,
thus calling the mathematical tractability into question.
However, like the non-single-letter fidelity criteria in 
\cite{BergerY1972,Kieffer1978b}, our fidelity criteria are 
tractable and also bear semantic significance on several applications.

All logarithms use base 2, and all rates are thus given in bits.
In this paper, rates are generally not normalized by the number of symbols $n$.
The reason for this departure from convention will soon become clear:
the total number of bits required in some problems scales sublinearly with $n$.
We use standard asymptotic notation such as 
$o(\cdot)$, $O(\cdot)$, $\Omega(\cdot)$ and
$\Theta(\cdot)$~\cite{OrlitskySVZ2006}.

\subsection{Outline}
The remainder of the paper is organized as follows. 
In Section~\ref{sec:ordervalue}, we propose the transformation of a
sequence into an order and a multiset of values, and we show that the order
and values are independent when letters are produced i.i.d.
Sections~\ref{sec:entropyrate} and~\ref{sec:universal} address lossless coding.
First we consider lossless coding with known distribution for both
finite and countably-infinite alphabets in Section~\ref{sec:entropyrate}.
Section~\ref{sec:universal} considers the universal setting and provides
both a positive result
(achievability of a rather low coding rate of 1 bit per letter) and
a negative result (unachievability of negligible redundancy).

In Sections~\ref{sec:lossy} and \ref{sec:universalLossy}, we turn to
lossy coding.
In Section~\ref{sec:RDzero} it is shown that, for a large class of sources,
a natural rate-distortion function is the trivial zero-zero point.
This inspires restriction to finite-sized blocks in
Sections~\ref{sec:finitefinite} and \ref{sec:finiteuncountable}.
These sections discuss the rate-distortion functions for the
discrete- and uncountable-alphabet cases, respectively.
Section~\ref{sec:finiteuncountable} also presents several
high-rate quantization analyses.
Universal lossy coding is considered in Section~\ref{sec:universalLossy}.
Finally, Section~\ref{sec:seq2set} concludes the paper with
a discussion of intermediates between full relevance and full irrelevance
of order;
additional connections to related work; and
a summary of the main results.

Many results presented here appeared first
in~\cite{VarshneyG2006b,VarshneyG2006,Varshney2006,VarshneyG2007}.

\section{Separating Order and Value}
\label{sec:ordervalue}
Consider source variables $X_1,\,X_2,\,\ldots,\,X_n$ drawn from
a common alphabet $\mathcal{X}$ according
to any joint distribution.
A realization $(x_i)_{i=1}^n$ can be decomposed into 
a multiset of values $\{x_i\}_{i=1}^n$ and an order $j$.  
This can be expressed as
\begin{equation}
\label{eq:transform}
(x_1,\,x_2,\,\ldots,\,x_n)
 = 
(y_{i_1},\,y_{i_2},\,\ldots,\,y_{i_n})
\longrightarrow
\left( {\begin{array}{*{20}c}
   {i_1 } & {i_2 } &  \cdots  & {i_n }  \\
   {x_1 } & {x_2 } &  \cdots  & {x_n }  \\
\end{array}} \right) = \left( {\begin{array}{*{20}c}
   {j}  \\
   {\left\{x_i\right\}_{i=1}^n}  \\
\end{array}} \right)
 \mbox{,}
\end{equation}
where $(y_i)_{i=1}^n$ is $(x_i)_{i=1}^n$ put into a canonical
order.\footnote{We could say that $y$ is the sorted version of $x$,
but we want to emphasize that there is no need at this point
for a \emph{meaningful} order for $\mathcal{X}$.}
The indices $i_1,\,i_2,\,\ldots,\,i_n$ are a permutation of the 
integers $1,\,2,\,\ldots,\,n$ and a deterministic function of $(x_i)_{i=1}^n$;
when the $x_i$s are not distinct, we require any deterministic
mechanism for choosing amongst the permutations such that
(\ref{eq:transform}) holds.
The ordering is collapsed into a single variable $j$,
which defines a chance variable $J$.  

The decomposition into order and value can be interpreted as the
generation of ``transform coefficients.''
Whether decomposing signals into low frequency and
high frequency~\cite{Dudley1939};
predictable and unpredictable~\cite{Oliver1952}; 
style and content~\cite{TenenbaumF2000}; object and texture~\cite{YanS1977};
or dictionary and pattern~\cite{OrlitskySZ2004}, 
divide-and-conquer approaches have been used to good advantage in many
source coding scenarios.
Here, we are concerned with situations in which the order $J$ is
irrelevant and hence allocated no bits.
In contrast, allocating all the bits to $J$ yields
permutation source codes~\cite{BergerJW1972}.
Other rate allocations are discussed in~\cite{VarshneyG2006b}.  

If the joint distribution of $(X_i)_{i=1}^n$ is exchangeable,
$J$ and $\left\{X_i\right\}_{i=1}^n$ 
are statistically independent chance variables \cite{JiTK1995}.
Thus they could be coded separately without loss of optimality. 
This is expressed in information theoretic terms by the following theorem.
\begin{thm}
\label{thm:Hdecomp}
For exchangeable sources,
the order and the value are independent,
and the sequence entropy $H((X_i)_{i=1}^n)$ can be decomposed 
into the value entropy $H(\{X_i\}_{i=1}^n)$ and the order entropy $H(J)$:
\begin{equation}
\label{eq:Hdecomp}
H\left( (X_i)_{i=1}^n \right) = H\left( \{X_i\}_{i=1}^n \right) + H(J) \mbox{.}
\end{equation}
\end{thm}
\begin{IEEEproof}
Suppressing unnecessary subscripts, we can write
\begin{align}
\label{eq:Hdecomp}
H((X)) & \stackrel{(a)}{=}  H((X)) + H(\{X\}) -  H((X)|J)
       \ = \                H(\{X\}) + I((X) \; ; \; J) \notag \\
       & =                  H(\{X\}) + H(J) - H(J|(X))
      \ \stackrel{(b)}{=} \ H(\{X\}) + H(J) \mbox{.} \notag
\end{align}
Step (a) follows from noting that $H(\{X\}) = H((X)|J)$ for exchangeable sources,
since all orderings are equiprobable and uninformative about the value. 
Step (b) follows from the fact $H(J|(X)) = 0$,
since the sequence determines the order. 
The other steps are simple informational manipulations.
\end{IEEEproof}

When we disregard order, we are just left with a multiset.
Type classes---also known variously as histograms or empirical distributions
in statistics and as rearrangement classes or Abelian classes
in combinatorics---are complete, minimal sufficient statistic for multisets.  
For discrete-alphabet sources,
types are convenient mathematical representations for multisets;
several results for these sources will depend
on counting numbers of type classes.

Despite being sufficient statistics,
types are not useful in the representation of continuous-alphabet sources.
As Csisz\'{a}r~\cite{Csiszar1998} writes: ``extensions of
the type concept to continuous alphabets are not known.''
In the case that $\mathcal{X} = \mathbb{R}$,
we make extensive use of 
the basic distribution theory of order statistics.
When the sequence of random variables $X_1,\ldots,X_n$
is arranged in ascending order as 
$X_{(1:n)} \le \cdots \le X_{(n:n)}$,
$X_{(r:n)}$ is called the $r$th order statistic.  It can be shown that 
the order statistics for exchangeable variates are complete, minimal 
sufficient statistics~\cite{DavidN2003}. 
For alphabets of vectors of real numbers,
there is no simple canonical form for expressing
the minimal sufficient statistic since there
is no natural ordering of vectors~\cite{Barnett1976}.

\section{Lossless Coding}
\label{sec:entropyrate}
Consider the lossless coding of multisets of $n$ letters
drawn from the discrete alphabet $\mathcal{X}$.
Since there are $n!$ permutations of a sequence of length $n$,
it would seem that a rate savings of $\log(n!)$ relative to sequence coding
might be possible.
Specifically,
the upper bound $H(J) \leq \log(n!)$ combined with
(\ref{eq:Hdecomp}) gives the lower bound
\begin{equation}
H(\{X_i\}_{i=1}^n) \ge H((X_i)_{i=1}^n) - \log n! \mbox{.}
\label{eq:orderlowerbound}
\end{equation}
Since this lower bound can be negative,
there must be more to the story. 

The lower bound is not tight due to the positive chance of ties among members
of a multiset drawn from a discrete parent.  If the chance of ties is small 
(if $|\mathcal{X}|$ is sufficiently large and $n$ is sufficiently small), 
the lower bound is a good approximation.

For any given source distribution and $n$,
a multiset of samples $\{x_i\}_{i=1}^n$ can be cast as 
a superletter drawn from an alphabet of multisets,
itself having a known distribution.
By the lossless block-to-variable source coding theorem~\cite{Shannon1948},
the entropy of the superletter is an asymptotically tight lower bound
on the rate required for representation.  
Since the type specifies the multiset,
as mentioned in Section~\ref{sec:ordervalue},
we can write
\begin{equation}
H(\{X_i\}_{i=1}^n) = H(K_1,K_2,\ldots,K_{|\mathcal{X}|}) \mbox{,}
\label{eq:multinomialEquiv}
\end{equation}
where $K_i$ is the number of occurrences of $x_i$ in $n$ trials.
While we will exhibit a few explicit calculations,
our main interest is in relating the rate requirement to the sample size $n$.
For this we first consider finite alphabets and then infinite alphabets.
  
\subsection{Finite Alphabets}
If the multiset is drawn i.i.d., the distribution of types is given 
by a multinomial distribution derived from the parent 
distribution \cite[Problem VI]{deMoivre1756}. 
Suppose $x_i \in \mathcal{X}$ has probability $p_i$ in the parent.
Then
\[
\Pr[K_i = k_i] = \binom{n}{k_1,k_2,\ldots,k_{|\mathcal{X}|}} \prod_{i=1}^{|\mathcal{X}|}p_i^{k_i}, \qquad \mbox{for $i = 1,\ldots, |\mathcal{X}|$,}
\]
for any type $(k_1,k_2,\ldots,k_{\left|\mathcal{X}\right|})$
of non-negative integers with sum $n$. 

The simplest case of a binary source ($|\mathcal{X}| = 2$)
gives $K_1 \sim \binomial(n,p)$ and $K_2 = n - K_1$, where
$p = \Pr[X_i = 1]$.
Then since $K_2$ is a deterministic function of $K_1$,
we have the simplification $H(K_1,K_2) = H(K_1)$.
Now we have
\begin{equation}
  H(\{X_i\}_{i=1}^n)
   = H(K_1)
   = \Half \log(2\pi e p(1-p) n) + \sum_{k=1}^\infty a_k n^{-k}
\label{eq:binomialEntropy}
\end{equation}
for some constants $a_1,\,a_2,\,\ldots$.
The leading term can be obtained with the de Moivre approximation
of a binomial random variable with a Gaussian random
variable~\cite[pp.\ 243--259]{deMoivre1756};
the full expansion requires more sophisticated techniques~\cite{JacquetS1999}.

To emphasize the dependence on $n$,
note that the rate in (\ref{eq:binomialEntropy}) is
$\Half \log n + c + o(1)$, where $c$ is some constant.
We will now see that $O(\log n)$ lossless coding rate
extends to all finite-alphabet multiset sources.

\begin{thm}
Let $|\mathcal{X}|$ be finite.
Then $H(\{X_i\}_{i=1}^n) = O(\log n)$.
\label{thm:HRzero}
\end{thm}
\begin{IEEEproof}
Denote the alphabet of distinct types by $\mathcal{K}(\mathcal{X},n)$.
By simple combinatorics~\cite{Csiszar1998},
\begin{equation}
|\mathcal{K}(\mathcal{X},n)| = \binom{n + |\mathcal{X}| - 1}{|\mathcal{X}| - 1} \le (n+1)^{|\mathcal{X}|} \mbox{.}
\label{eq:NofTypes}
\end{equation}
Recalling the equality (\ref{eq:multinomialEquiv}),
the desired entropy $H(\{X_i\}_{i=1}^n)$
is upper-bounded by the logarithm of
$|\mathcal{K}(\mathcal{X},n)|$.
Thus,
$$
  H(\{X_i\}_{i=1}^n) \leq |\mathcal{X}| \log(n+1)
                     = O(\log n)
$$
since $|\mathcal{X}|$ is finite.
\end{IEEEproof}

Note that the theorem holds for any source, not just for i.i.d.\ sources.
For a non-trivial i.i.d.\ source we can use the calculation
(\ref{eq:binomialEntropy}) to show an $\Omega(\log n)$ lower bound,
so in fact we have $H(\{X_i\}_{i=1}^n) = \Theta(\log n)$.

For an i.i.d.\ source,
the upper bound in the proof is quite loose.  
To achieve the bound with equality,
each of the types would have to be equiprobable; however by 
the strong asymptotic equipartition property~\cite{Yeung2002}, collectively,
all non-strongly typical types will occur with arbitrarily small probability.
The number of types in 
the strongly typical set is polynomial in $n$,
so any upper bound would still be $\Theta(\log n)$.

\subsection{Countable Alphabets}
\label{sec:losslessCountable}
Theorem~\ref{thm:HRzero} with its presented proof obviously does not extend
to infinite alphabets.
To get an interesting bound we must do more than enumerate types.

Define the entropy rate of a multiset as 
\begin{equation}
H(\mathfrak{X}) = \lim_{n \rightarrow \infty} \tfrac{1}{n}H(\{X_i\}_{i=1}^n) \mbox{.}
\end{equation}
Theorem~\ref{thm:HRzero} shows that finite-alphabet sources yield
multisets with zero entropy rate.
Using a dictionary--pattern decomposition,
we will show a related result for countable-alphabet sources.

A sequence may be decomposed into a \emph{dictionary}, $\Delta$,
and a \emph{pattern}, $(\Psi_i)$,
where the dictionary specifies which letters from the 
alphabet have appeared and the pattern specifies the order in which
these letters have appeared \cite{OrlitskySZ2004}.  For a sequence
$(x_i)$, dictionary entry $\delta_k \in \mathcal{X}$ is the $k$th 
distinct letter in the sequence and pattern entry $\psi_i \in \mathbb{Z}^{+}$
is the dictionary index of the $i$th letter 
in the sequence.  Note that the type of a pattern, denoted 
as $\{\Psi_i\}$, and the pattern of a multiset, denoted as 
$\Psi(\{X_i\})$, are the same.  It can be seen that a multiset is 
determined by $\Delta$ and $\{\Psi_i\}$; the order of the 
pattern, $J(\Psi)$, is not needed.   

Based on \cite{OrlitskySVZ2006}, we show that the entropy rate 
of a multiset generated by a discrete finite-entropy stationary 
process and the entropy rate of its pattern are equal.  
First note that the dictionary of a sequence and the dictionary
of its associated multiset are the same; we use $\Delta$ to signify 
either one.  Since $\{X_i\}$ determines $\Psi(\{X_i\})$ and since 
given $\{\Psi_i\}$, there is a one-to-one correspondence between 
$\{X_i\}$ and $\Delta$, 
\[
H(\{X_i\}) = H(\{\Psi_i\}) + H(\{X_i\} \mid \{\Psi_i\}) = H(\{\Psi_i\}) + H(\Delta \mid \{\Psi_i\}) \mbox{.}
\]
If we can show that $H(\Delta | \{\Psi_i\})$ is $o(n)$, then it will follow 
that the entropy rate of the multiset, $H(\mathfrak{X})$, is equal to 
the entropy rate of the pattern of the multiset
\[
H(\mbox{\Neptune}) = \lim_{n\to\infty}\tfrac{1}{n} H(\{\Psi_i\}_{i=1}^n) \mbox{.}
\]

Noting the fact that the dictionary $\Delta$ is independent 
of the order of the pattern, $J(\Psi)$, we establish that
\begin{align*}
\lim_{n\to\infty}\tfrac{1}{n} H(\Delta|(\Psi_i)_{i=1}^n) &= \lim_{n\to\infty}\tfrac{1}{n} H(\Delta | J((\Psi_i)_{i=1}^n),\{\Psi_i\}_{i=1}^n) \\                            
&= \lim_{n\to\infty}\tfrac{1}{n} H(\Delta \mid \{\Psi_i\}_{i=1}^n) \mbox{,}
\end{align*}
where the first step follows since the order and type of pattern determine 
the pattern, and the second step is due to independence.  The result 
\cite[Theorem 9]{OrlitskySVZ2006} shows that for all 
finite-entropy, discrete stationary processes, the asymptotic per-letter 
values of $H((\Psi_i)))$ and $H((X_i))$ are equal.  Hence,
\[
\lim_{n\to\infty}\tfrac{1}{n} H(\Delta \mid (\Psi_i)_{i=1}^n) = 0 \mbox{,}
\]
and so
\[
\lim_{n\to\infty}\tfrac{1}{n} H(\Delta \mid \{\Psi_i\}_{i=1}^n) = 0 \mbox{.}
\]
This implies that $H(\Delta | \{\Psi_i\})$ is $o(n)$ and yields the following theorem.
\begin{thm}   
\label{thm:coincide}
The entropy rate of a multiset generated by a discrete 
finite-entropy stationary process and the entropy rate of 
its pattern coincide:
\[ 
H(\mathfrak{X}) = H(\mbox{\Neptune}) \mbox{.}
\]
\end{thm}

Computing the entropy rate of the multiset or equivalently of 
the pattern of the multiset can be difficult.  See \cite{Shamir2007} and 
references therein for a discussion on computing the entropy of patterns;
the entropy computation for patterns of multisets is closely related.

\section{Universal Lossless Coding}
\label{sec:universal}
The previous section considered source coding for multisets when the 
source distribution was known.  Most prominently
in estimation and inference but also in other applications,
the underlying distribution is not known.  In this section, we discuss universal
source coding of multisets, first presenting an achievability result
for countable alphabets and then showing that redundancy, defined in a stronger sense
than usual, cannot be driven to zero.

\subsection{Universal Achievability}
\label{sec:achievability}
We propose a source coding scheme that
achieves some degree of compression for all members of a source class
at the same time.  We will not compare to the entropy bound, 
holding off detailed discussion of redundancy until Section~\ref{sec:Conv}.

Consider classes of countable-alphabet i.i.d.\ sources that meet 
Kieffer's condition for universal encodability for the sequence representation
problem \cite{Kieffer1978,HeY2005}.  
For these source classes,
the redundancy for encoding $(X_i)_{i=1}^n$ is $o(n)$. 
We formulate a universal scheme for the multiset representation 
problem and demonstrate an achievability result, making use
of the dictionary--pattern decomposition.

As we saw in Section~\ref{sec:losslessCountable}, a multiset can be represented
as the concatenation of the pattern of the multiset and the dictionary. 
Consider the rate requirements of these
two parts separately.
First, let us bound the rate that is required to represent the
pattern of the multiset (the type of the pattern).
We can make use of the fact that there are $2^{n-1}$ types of 
patterns.  This enumeration follows because the types are 
sequences of positive integers that sum to $n$.  These can appear
in any order, thus we are counting ordered partitions.  It is well
known that there are $2^{n-1}$ ordered partitions, which can be
seen as determining arrangements of $n-1$ possible separations 
of $n$ places.  Thus the rate requirement for an enumerative universal scheme
representing the type of the pattern is $n-1$ bits.

Now to determine the rate requirement of the dictionary given the type 
of pattern.  If the underlying distribution were known, we saw 
in Theorem \ref{thm:coincide} that $H(\Delta | \{\Psi_i\}_{i=1}^n)$ is $o(n)$.  
It was shown in \cite{OrlitskySZ2004} that there is
an $O(\sqrt{n})$ upper bound on the pattern
redundancy, independent of $|\mathcal{X}|$. 
Since this is sublinear,  
the asymptotic per-letter redundancy in coding a class of 
sequences $(X_i)_{i=1}^{n}$ from a countable alphabet
coincides with the asymptotic per-letter redundancy in 
coding the dictionary given the pattern.  Since we are
considering a class that meets Kieffer's condition, we find
the redundancy in coding the dictionary given the pattern
is $o(n)$.  This also carries over to coding the dictionary
given the pattern of the multiset, due to the independence
between dictionary and order of pattern that we had put forth
in Section~\ref{sec:losslessCountable}.  Since both $H(\Delta|\{\Psi_i\}_{i=1}^n)$
and the redundancy in coding the dictionary given the pattern
of the multiset are $o(n)$, the total rate requirement
for coding the dictionary given the pattern of the multiset
is $o(n)$.

Adding together the rate requirements for the two parts
yields the following achievability theorem.
The rate requirement is universally reduced from $[0,\infty)$ 
bits per letter for the sequence problem to $1$ bit per letter for 
the multiset problem.
\begin{thm}
\label{thm:universalAchievability}
Given any i.i.d.\ source class that is universally
encodable as a sequence, the multiset $\{X_i\}_{i=1}^n$ can be encoded
with $n + o(n)$ bits.
\end{thm}
\begin{IEEEproof}
A representation consists of the concatenation of the type of pattern 
and the dictionary given the type of pattern.  The first part requires
$n-1$ bits.  The second part requires $o(n)$ bits.  The total rate
is then $n + o(n)$.
\end{IEEEproof}

\begin{figure}
  \centering
  {
   \psfrag{X}[][]{$\mathcal{X}$}
   \psfrag{Y}[][]{}
   \includegraphics[width=2.2in]{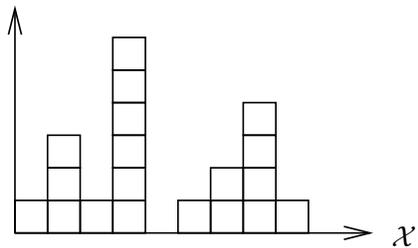}
  }
  \caption{A histogram for interpretation of
   Theorem~\ref{thm:universalAchievability}.  Starting at the left and
   using the rule described in the text, the histogram is encoded by
   {\tt 01001100000111010001}.}
  \label{fig:histogram}
\end{figure}

Coding a multiset is equivalent to coding a type or histogram.
An interpretation of Theorem~\ref{thm:universalAchievability} is
thus that histograms, from a certain class, with total weight $n$ can be
encoded with $n + o(n)$ bits.
Figure~\ref{fig:histogram} shows a histogram.
An encoding method that shows the plausibility of $n + o(n)$ total rate
is to encode the histogram one letter at a time,
starting from the left end and moving to the right.
After each letter, use {\tt 0} to indicate that there is another
occurrence of the same letter and use {\tt 1} to move on to the next letter.
If every symbol in the alphabet appears at least once, the rate is $n$;
as long as the right-most letter encountered does not grow too quickly with $n$,
an $n + o(n)$ rate is achieved.

\subsection{Unattainability of Negligible Redundancy}
\label{sec:Conv}

For finite alphabets, we showed that the multiset entropy rate is zero 
for any source; this is a crude consequence of Theorem~\ref{thm:HRzero}.
We also saw, in the proof of Theorem~\ref{thm:HRzero},
that simply enumerating the type classes
requires zero rate per multiset letter asymptotically. 
Hence such a universal scheme requires zero rate for any finite-alphabet source.
However, we cannot conclude from ``zero equals zero'' that the excess
rate due to not knowing the source distribution is negligible.

In this section we take a finer look at universal lossless coding of multisets.
We will see that for a finite-alphabet source,
the redundancy cannot be made a negligible fraction of the coding rate.
Since the coding rate with full distributional knowledge is $\Theta(\log n)$,
we come to this conclusion using new information and redundancy measures
that have normalization by $\log n$ rather than by $n$.
We will find that zero-redundancy universal coding of multisets 
is \emph{not} possible with respect to the class of
memoryless multisets, using the more stringent redundancy definition.
Zero redundancy is thus also not possible for more general classes of sources
such as sources with memory or with infinite alphabets.

\subsubsection{Log-Blocklength Normalized Information Measures}
We formulate several definitions and extend 
the source coding theorems to these definitions.  
Let $Z_1^1,\,Z_1^2,\,\ldots$ represent a sequence
of random variables over a sequence of alphabets
$\mathcal{Z}_1,\,\mathcal{Z}_2,\,\ldots$.
(For sequence coding we would have $Z_1^n = (X_i)_{i=1}^n$
and $\mathcal{Z}_n = \mathcal{X}^n$ is an alphabet of sequences.
For multiset coding we would have $Z_1^n = \{X_i\}_{i=1}^n$
and $\mathcal{Z}_n = \mathcal{K}(\mathcal{X},n)$ is an alphabet of types.)
Define the log-blocklength normalized entropy rate as
\[
\mathfrak{H}(\mathfrak{Z}) = \lim_{n \to \infty} \frac{H(Z_1^n)}{\log n}
\]
when the limit exists. 
With conditioning on another random variable $\Theta$,
define the log-blocklength normalized conditional entropy rate as
\[
\mathfrak{H}(\mathfrak{Z} \mid \Theta) = \lim_{n \to \infty} \frac{H(Z_1^n \mid \Theta)}{\log n}
\]
when the limit exists.  Similarly, define the log-blocklength normalized 
information rate as
\[
\mathfrak{I}(\mathfrak{Z};\Theta) = \lim_{n \to \infty} \frac{I(Z_1^n;\Theta)}{\log n}
\]
when the limit exists.
While these definitions parallel the standard definitions,
none of the limits would generally exist for sequence coding
because the numerators grow linearly with $n$.

For each $n \in \mathbb{Z}^+$, let $\phi_n$ be a source code for
random variable $Z_1^n$.
For this sequence of source codes, the average codeword lengths are
\[
C_{\phi,n} = \sum_{\mathcal{Z}_n} p_{Z_1^n}(z_1^n) \ell(z_1^n) \mbox{,}
\]
where $\ell(\cdot)$ is the length of the codeword assigned 
to the source realization $z_1^n$.
Shannon's fixed-to-variable 
source coding theorem \cite{Shannon1948} establishes that 
there exists a sequence of source codes that satisfy the following inequalities
for all $n$:
\[
H(Z_1^n) \le C_{\phi,n} \le H(Z_1^n) + 1 \mbox{.}
\]
Dividing through by $\log n$ yields
\[
\frac{H(Z_1^n)}{\log n} \le \frac{C_{\phi,n}}{\log n} \le \frac{H(Z_1^n) + 1}{\log n} \mbox{.}
\]
Taking the limit of large blocklength ($n \rightarrow \infty$),
we see that when the limits exist there is a sequence of 
source codes that achieves $\mathfrak{H}(\mathfrak{Z})$.

\subsubsection{Log-Blocklength Normalized Redundancy Measures}
Define the redundancy of a source code, $r_{\phi,n}$, as 
the excess average codeword length that is required over 
the minimum $H(Z_1^n)$:
\[
r_{\phi,n} = C_{\phi,n} - H(Z_1^n) \mbox{.}
\]
Finally, define the log-blocklength normalized redundancy 
of a sequence of source codes as
\[
\lim_{n\to\infty}\frac{r_{\phi,n}}{\log n} = \lim_{n\to\infty}\frac{C_{\phi,n} - H(Z_1^n)}{\log n} 
= \lim_{n\to\infty}\frac{C_{\phi,n}}{\log n} - \mathfrak{H}(\mathfrak{Z}) \stackrel{\Delta}{=} \mathfrak{C}_{\phi} - \mathfrak{H}(\mathfrak{Z}) \mbox{.}
\]
By the manipulations of the source coding theorem that we had 
made previously, 
we know that there is a sequence of codes with $\mathfrak{C} = 0$.  The 
code used to develop
the upper bound in the source coding theorem, however, requires 
that $p_{Z_1^n}(z_1^n)$ is 
known. 

Now we define performance measures 
for source coding for a class of 
source distributions, rather than just a single source distribution.  
The definitions parallel those of \cite{Davisson1973}.  
Suppose that the source distribution is chosen from a class that 
is parameterized by 
$\Theta \in \mathcal{T}$.  For each $\theta$, there is a 
conditional distribution
\[
p(z_1^n \mid \theta) = \Pr\left[Z_1^n = z_1^n \mid \Theta = \theta\right] \mbox{.}
\]
The parameter $\theta$ is fixed but unknown, when generating 
the source realization.  
Moreover, there may be a distribution on this 
parameter, $p_{\Theta}(\theta)$. 
Let $\Phi_n$ be the set of all uniquely decipherable codes on $Z_1^n$.
Then, the 
average log-blocklength normalized redundancy of a 
code $\phi \in \Phi_n$ for the class of 
sources described by $p_{\Theta}(\theta)$ is
\[
\mathcal{L}_{\phi,n}(p_{\Theta}) = \int_{\mathcal{T}} \frac{r_{\phi,n}}{\log n} p_{\Theta}(\theta) \, d\theta \mbox{.}
\]

The minimum 
$n$th-order average log-blocklength normalized 
redundancy is
\[
\mathcal{L}_{n}^{*}(p_{\Theta}) = \inf_{\psi \in \Phi_n} \mathcal{L}_{\psi,n}(p_{\Theta}) \mbox{.}
\]
Finally, the minimum average log-blocklength normalized 
redundancy is
\[
\mathcal{L}^{*}(p_{\Theta}) = \lim_{n\to\infty} \mathcal{L}_{n}^{*}(p_{\Theta}).
\]
If $\mathcal{L}^{*}(p_{\Theta}) = 0$, then a sequence of codes that 
achieves the limit is called
\emph{weighted log-blocklength normalized universal}.
Now let $T$ be the set of all probability distributions defined 
on the alphabet $\mathcal{T}$.  Then the 
$n$th-order maximin log-blocklength normalized redundancy 
of $T$ is
\[
\mathcal{L}_{n}^{-} = \sup_{q_{\Theta} \in T} \mathcal{L}_{n}^{*}(q_{\Theta}) \mbox{.}
\]
If it exists, then the maximin log-blocklength normalized redundancy is
\[
\mathcal{L}^{-} = \lim_{n\to\infty}\mathcal{L}_{n}^{-} \mbox{.}
\]
If $\mathcal{L}^{-} = 0$, then a sequence of codes 
that achieves the limit is called
\emph{maximin log-blocklength normalized universal}.
The $n$th-order minimax log-blocklength normalized 
redundancy of $\mathcal{T}$ is 
\[
\mathcal{L}_{n}^{+} = \inf_{\phi\in\Phi_n}\sup_{\theta\in\mathcal{T}} \frac{r_{\phi,n}(\theta)}{\log n}
\]
and the minimax log-blocklength normalized redundancy 
of $\mathcal{T}$ is
\[
\mathcal{L}^{+} = \lim_{n\to\infty}\mathcal{L}_{n}^{+} \mbox{.}
\]
If $\mathcal{L}^{+} = 0$, then a sequence of codes 
that achieves the limit is called
\emph{minimax log-blocklength normalized universal}.

\subsubsection{Redundancy-Capacity Theorems}
The senses of universality that we have defined obey an ordering relation.
\begin{thm} 
\label{thm:relation}
The log-normalized redundancy quantities satisfy
\[
\mathcal{L}_n^{+} \ge \mathcal{L}_n^{-} \ge \mathcal{L}_n^{*}(p_{\Theta})
\]
and 
\[
\mathcal{L}^{+} \ge \mathcal{L}^{-} \ge \mathcal{L}^{*}(p_{\Theta}) \mbox{.}
\]
\end{thm}
\begin{IEEEproof}
Minor modification of \cite[Theorem 1]{Davisson1973}.
\end{IEEEproof}

Armed with definitions and relations among several notions 
of log-blocklength normalized universality, we now study when 
it is possible to achieve universality.  
We give a theorem that gives a necessary and 
sufficient condition on the existence of 
weighted log-blocklength normalized universal 
codes.  
\begin{thm}
\label{thm:MIuniversal}
The minimum $n$th-order average log-blocklength 
normalized redundancy is bounded as
\[
\frac{I(Z_1^n;\Theta)}{\log n} \le \mathcal{L}_{n}^{*}(p_{\Theta}) \le \frac{I(Z_1^n;\Theta)}{\log n} + \frac{1}{\log n} \mbox{.}
\]
A necessary and sufficient condition for the existence of 
weighted log-blocklength normalized universal
codes is that 
\[
\mathcal{L}^{*}(p_{\Theta}) = \lim_{n\to\infty} \mathcal{L}_{n}^{*}(p_{\Theta}) = \lim_{n\to\infty} \frac{I(Z_1^n;\Theta)}{\log n} = \mathfrak{I}(Z;\Theta) = 0 \mbox{.}
\]
\end{thm}
\begin{IEEEproof}
Minor modification of \cite[Theorem 2]{Davisson1973}.
\end{IEEEproof}

Theorem \ref{thm:MIuniversal} can be extended to conditions for minimax and maximin 
log-blocklength normalized universality and can also be strengthened 
by suitable modification of theorems in \cite{Gallager1979,MerhavF1995}.

\subsubsection{Class of Memoryless Multisets}
\label{sec:universalMemorylessMultisets}
Consider the class of memoryless, binary multisets.  The 
parameter $\theta$ is the Bernoulli trial parameter.  
Now suppose that there is a distribution over the 
parameter space $q_{\Theta}(\theta) \in T$ 
that is uniform over $[0,1]$.  This gives a mixed 
source where all type classes are equiprobable,
as shown now. 

Let the realizations of $\{X_i\}_{i=1}^n$ 
be expressed as $z \in \{0,\ldots,n\}$,
the number of ones.  
\[
\Pr[\{X_i\}_{i=1}^n = z] = \int_{0}^{1} q_{\Theta}(\theta) \Pr[\{X_i\}_{i=1}^n = z \mid \Theta = \theta] \, d\theta \mbox{,}
\]
where $q_{\Theta}(\theta)$ simply equals one over the range of 
integration, and $\Pr[\{X_i\}_{i=1}^n = z \mid \Theta = \theta]$
is given by a binomial distribution:
\[
\Pr[\{X_i\}_{i=1}^n = z \mid \Theta = \theta] = \binom{n}{z}\theta^z (1-\theta)^{n-z} \mbox{.}
\]
So,
\[
\Pr[\{X_i\}_{i=1}^n = z] = \int_{0}^{1} \binom{n}{z}\theta^z (1-\theta)^{n-z} \, d\theta = \binom{n}{z} B(z+1,n-z+1) = \frac{1}{1+n} \mbox{,}
\]
where the beta function $B(\cdot,\cdot)$ has been used.  Thus, the 
result is that the type classes are equiprobable.

Since the types are equiprobable, the entropy
$H(\{X_i\}_{i=1}^n)$ is just the logarithm of the number of the types:
\[
H(\{X_i\}_{i=1}^n) = \log \binom{n+|\mathcal{X}|-1}{|\mathcal{X}|-1} = \log(n+1) \mbox{.}
\]
The entropy conditioned on $\Theta = \theta$ is simply the entropy
of a binomial random variable \cite{JacquetS1999} and so the conditional entropy is 
\begin{align*}
H(\{X_i\}_{i=1}^n \mid \Theta)
 &= \int H(\{X_i\}_{i=1}^n \mid \theta) q_{\Theta}(\theta) \, d\theta
  = \int \left[\tfrac{1}{2}\log\left(2\pi e n \theta(1-\theta)\right) + {\ts\sum_{k\ge 1}a_k n^{-k}}\right]q_{\Theta}(\theta) \, d\theta \\
 &= \tfrac{1}{2}\log n + \sum_{k\ge 1}n^{-k} \int a_k q_{\Theta}(\theta) \, d\theta  + \int \tfrac{1}{2}\log\left(2\pi e \theta(1-\theta)\right) q_{\Theta}(\theta) \, d\theta \\
 &= \tfrac{1}{2}\log n + o(\log n)\mbox{,}
\end{align*}
where the $a_k$s are known constants given in \cite{JacquetS1999}.

Now the mutual information is
$I(\{X_i\}_{i=1}^n;\Theta) = H(\{X_i\}_{i=1}^n) - H(\{X_i\}_{i=1}^n \mid \Theta)$,
so the log-blocklength
normalized information rate is given by
\begin{align*}
\mathfrak{I}
  &= \lim_{n\to\infty} \frac{I(\{X_i\}_{i=1}^n;\Theta)}{\log n}
   = \lim_{n\to\infty} \frac{\log(n+1) - \tfrac{1}{2}\log n - o(\log n)}{\log n} \\
  &= \lim_{n\to\infty} \frac{\log(n+1) - \tfrac{1}{2}\log n}{\log n}
   = \lim_{n\to\infty} -\frac{1}{2} + \frac{\log(n+1)}{\log n}
   = \frac{1}{2}.
\end{align*}
Since this is greater than $0$, we have shown that the weakest form of 
universality is not possible, by Theorem \ref{thm:MIuniversal}.  
Thus by Theorem \ref{thm:relation}, stronger forms of universality are 
not possible either.  Since the class of binary memoryless sets 
is a subset of more general source classes such as 
memoryless; Markov; and stationary, ergodic,
universal source coding over these source classes is not possible either. 

We can calculate the weighted redundancy for classes of memoryless sources 
with larger alphabet sizes.  Using the $p_{\Theta}(\theta)$ that yields
equiprobable multisets and the conditional entropy given by the
entropy of a multinomial random variable \cite{JacquetS1999}, we are interested in
\[
\lim_{n\to\infty} \frac{\log\binom{n+|\mathcal{X}|-1}{|\mathcal{X}|-1} - \Half(|\mathcal{X}|-1)\log(K n) - o(1)}{\log n} = \frac{|\mathcal{X}|-1}{2} \mbox{,}
\]
where $K$ is a known constant.
As we can see, this redundancy grows without
bound as the alphabet size increases.  Perhaps unsurprisingly, this redundancy 
expression is reminiscent of the unnormalized redundancy expression 
for i.i.d.\ sequences \cite{DrmotaS2004}:
\[
\frac{|\mathcal{X}|-1}{2} \log\frac{n}{2\pi} + \log\frac{\Gamma^{|\mathcal{X}|}\left(1/2\right)}{\Gamma\left({|\mathcal{X}|}/{2}\right)} + o_{|\mathcal{X}|}(1) \mbox{.}
\]
We see that the richness of a class of sources as sequences is the same
as the richness of that class as multisets.

Let us also comment that in the deterministic case of individual multisets, 
rather than the probabilistic classes of sources that we have been considering,
the same non-achievability result applies.  This follows from the arguments 
summarized in \cite{MerhavF1998}.

\section{Lossy Coding}
\label{sec:lossy}
In the two previous sections, we have considered lossless representation 
of multiset sources with discrete alphabets.  Now in this section and
Section~\ref{sec:universalLossy},
we look at lossy coding with both discrete and continuous alphabets.

\subsection{Large-Size Multiset Asymptotics}
\label{sec:RDzero}

\subsubsection{Multiset Mean Squared Error}
Assume that the source alphabet $\mathcal{X}$ is a subset of the real numbers.
In cases of interest---for example, when every $(X_i)_{i=1}^n$ has a
probability density---multisets drawn from these alphabets are almost
surely sets.  (Simply, ties have zero probability.)
Thus, the type is a list of $n$ values that each occurred once.
This list is conveniently represented with order statistics.
Recall that $X_{(r:n)}$ denotes the $r$th order statistic from a block of $n$,
which is the $r$th-largest of the set $\{X_i\}_{i=1}^n$.

Define a word distortion measure as
\begin{equation}
\label{eq:rhodef}
\rho_n(x_1^n,y_1^n) = \frac{1}{n}\sum_{i = 1}^n (x_{(i:n)} - y_{(i:n)})^2 \mbox{,}
\end{equation}
and an associated fidelity criterion as
\begin{equation}
\label{eq:F1defA}
F_1 = \{\rho_n(x_1^n,y_1^n), n=1,2,\ldots\}\mbox{.}
\end{equation}
Although not a single-letter fidelity criterion, it is single-letter 
mean square error on the block of order statistics.  

\begin{figure}
  \centering
  \includegraphics[width=3.0in]{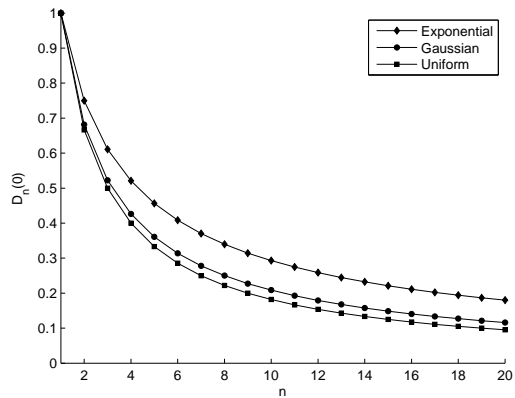}
  \caption{Distortion at zero rate $D_n(0)$ as function of multiset size $n$ for
   several sources.}
  \label{fig:AvgV}
\end{figure}

If $\{X_i\}_{i=1}^n$ is reconstructed by $\{Y_i\}_{i=1}^n$,
the incurred distortion is
\[
D_n = \frac{1}{n} \sum_{i=1}^n \E\left[\left(X_{(i:n)} - Y_{(i:n)}\right)^2\right] \mbox{.}
\]
If we use no rate, then the best choice for the reconstruction is simply 
$y_{(i:n)} = \E[X_{(i:n)}]$, $i=1,\,2,\,\ldots,\,n$,
and the average incurred distortion reduces to
\[
D_n(R=0) = \frac{1}{n}\sum_{i=1}^n \var\left(X_{(i:n)}\right) \mbox{.}
\]
Before proceeding with a general proof that $D(R=0)=\lim_{n\to\infty}D_n(R=0) = 0$, 
we give some examples.  For multisets with elements drawn i.i.d.\ from 
the uniform distribution with support $\left[-\sqrt{3},\sqrt{3}\right]$, 
from the Gaussian distribution with mean zero and variance one, and from 
the exponential distribution with mean one, Figure \ref{fig:AvgV} shows the
$n$th-order distortion-rate function. 
This is the average variance of the order statistics.
It can be shown that all of these bounded, monotonically decreasing sequences
of real numbers, $\{D_n(0)\}$, have limit $0$ \cite{Varshney2006}.  
In fact all of these sequences decay as $\Theta(1/n)$.  Hence for zero rate, 
there is zero distortion incurred.  

The result that the rate-distortion function is the zero-zero point, along
with the distortion decay as a function of block size being $\Theta(1/n)$,
also holds for a large class of other sources.  If we assume that 
the cumulative distribution function of the source is always differentiable
(i.e.\ the density function $p_X(x)$ exists) and that $p_X(x) > 0$ for all
$x \in \mathbb{R}$, then the same result holds.  This 
follows from the asymptotic fixed variance normality of $\sqrt{n}$-normalized central 
order statistics \cite[Corollary 21.5]{Vaart1998}.
Notice that although this class of sources is very large, 
two of our examples were not members.  Thus, we formulate
an even more widely-applicable theorem on zero rate-zero distortion,
though we no longer have a characterization of the decay rate.

The general theorem will be based on the quantile function of the 
i.i.d.\ parent process; this is the 
generalized inverse of the cumulative distribution function, $F_X(x)$,
\[
Q(w) = F^{-1}(w) = \inf\{x:F_X(x)\ge w \} \mbox{.}
\]
The empirical quantile function, defined in terms of order statistics is
\[
Q_n(w) = X_{(\lfloor wn\rfloor + 1:n)} = F_n^{-1}(w) \mbox{,}
\]
where $F_n(\cdot)$ is the empirical distribution function.  The quantile function $Q(\cdot)$ is continuous
if and only if the distribution function has no flat portions in the interior.  
The main step of the proof will be
a Glivenko-Cantelli like theorem for empirical quantile functions \cite{Mason1982}.

\begin{lemma}
\label{lemma:mason}
Let the letters to be coded, $X_1,\,X_2,\,\ldots,\,X_k$, be generated in an i.i.d.\ fashion according to
$F_X(x)$ with associated quantile function $Q(w)$.  Let $X_1$ satisfy
\begin{equation}
\E\left[|\min(X_1,0)|^{1/\nu_1}\right] < \infty \quad \mbox{and} \quad \E\left[(\max(X_1,0))^{1/\nu_2}\right] < \infty
\label{eq:bndmom}
\end{equation} 
for some $\nu_1 > 0$ and $\nu_2 > 0$
and have continuous quantile function $Q(w)$.
Then the sequence of distortion-rate values
for the coding of size-$n$ sets
drawn from the parent distribution satisfy
\[
\lim_{n \to \infty} D_n(R=0) = 0.
\]
\end{lemma}
\begin{IEEEproof}
For any nonnegative function $\omega$ defined on $(0,1)$, define a weighted Kolmogorov-Smirnov like statistic
\[
S_n(\omega) = \sup_{0 < w < 1} \omega(w)\left|Q_n(w) - Q(w)\right| \mbox{.}
\]
For each $\nu_1 > 0$, $\nu_2 > 0$, and $w \in (0,1)$, define the weight function
\[
\omega_{\nu_1,\nu_2}(w) = w^{\nu_1}(1-w)^{\nu_2} \mbox{.}
\]
Assume that $Q$ is continuous, choose any $\nu_1 > 0$ and $\nu_2 > 0$, and define 
\[
\gamma = \limsup_{n \to \infty} S_n(\omega_{\nu_1,\nu_2}) \mbox{.}
\]
Then by a result of Mason \cite{Mason1982}, $\gamma = 0$ with probability $1$ when (\ref{eq:bndmom}) holds.
Our assumptions on the parent process meet this condition, so $\gamma = 0$ with probability $1$.  This implies that 
\begin{equation}
\label{eq:convergenceOfOS}
\limsup_{n \to \infty } |X_{(\lfloor wn \rfloor  + 1:n)}  - Q(w)| \le 0 \mbox{ for all } w \in (0,1) \mbox{ w.p.} 1 \mbox{,}
\end{equation}
and since the absolute value is nonnegative, the inequality holds with equality.  According to (\ref{eq:convergenceOfOS}),
for sufficiently large $n$, each order statistic takes a fixed value with probability 1.  
The bounded moment condition on the parent process, (\ref{eq:bndmom}), implies
a bounded moment condition on the order statistics.  Almost sure convergence to a fixed quantity, together with
the bounded moment condition on the events of probability zero imply convergence in second moment of all order statistics.  
This convergence in second moment to a deterministic distribution implies that the variance of each order statistic is zero, 
and thus the average variance is zero.  
\end{IEEEproof}

We have established that asymptotically in $n$, the point $(R=0,D=0)$ is achievable, which leads
to the following theorem.
\begin{thm}
\label{thm:zerozero}
Under fidelity criterion $F_1$, $R(D) = 0$ for an i.i.d.\ 
source that meets the bounded moment condition (\ref{eq:bndmom}) and 
has continuous quantile function.
\end{thm}
\begin{IEEEproof}
By the nonnegativity of the distortion function, $D(R) \ge 0$. By Lemma~\ref{lemma:mason}, $D(0) \le 0$, 
so $D(0) = 0$.  Since $D(R)$ is a non-increasing function, $D(R) = 0$, and so $R(D) = 0$.
\end{IEEEproof}
Due to the generality of the Glivenko-Cantelli like theorem that we used, the result will stand 
for a very large class of distortion measures.  One only needs to ensure that the set of outcomes of 
probability zero is not problematic.

\subsubsection{Arbitrary Single-Letter Distortions}
\label{sec:lossyArbitrary}
While Theorem~\ref{thm:zerozero} applies to a large class of
real-value parent distributions, it depends on the multiset squared error
distortion measure (\ref{eq:rhodef}) to make convergence of moments of
the order statistics relevant.
With arbitrary single-letter distortion measures we
obtain a result analogous to Theorem~\ref{thm:HRzero}
in that it shows that an $O(\log n)$ rate is sufficient for
coding $n$ letters accurately.

Consider a source and single-letter distortion function
$d: \mathcal{X} \times \hat{\mathcal{X}} \rightarrow \mathbb{R}^+$
such that the rate distortion function (for encoding as a \emph{sequence})
is $R_X(D)$.
For coding this source without regard to order,
define a word distortion measure
\begin{equation}
\label{eq:rhodefGeneral}
  \rho_n(x_1^n,y_1^n) = \min_{\pi} \frac{1}{n}\sum_{i=1}^n d(x_i,y_{\pi(i)}),
\end{equation}
where $\pi$ is a permutation on $\{1,\,2,\,\ldots,\,n\}$,
and an associated fidelity criterion
\begin{equation}
\label{eq:F1defB}
  F_1 = \{\rho_n(x_1^n,y_1^n), n=1,2,\ldots\}\mbox{.}
\end{equation}
(We have re-used the notation from (\ref{eq:rhodef})--(\ref{eq:F1defA})
since the per-letter MSE on order statistics defined there is a special case.)
Denote the minimum (total) rate for encoding $\{X_i\}_{i=1}^n$ with
$\E[\rho_n(X_1^n,\hat{X}_1^n)] \leq D$ by
$R_{ \{X_i\}_{i=1}^n }(D)$.
Then the following theorem bounds the growth of 
$R_{ \{X_i\}_{i=1}^n }(D)$ as a function of $n$.

\begin{thm}
\label{thm:lossy_logn}
If $R_X(D)$ is finite, then for any $\epsilon > 0$, 
\[
R_{\{X_i\}_{i=1}^n}(D + \epsilon) = O(\log n) \mbox{.}
\] 
\end{thm}
\begin{IEEEproof}
Let $D$ be such that $R = R_X(D)$ is finite and let $\epsilon > 0$.
The achievability of $R_X(D)$ means there is sequence of dimension-$n$
quantizers with $2^{nR}$ codewords such that
$\lim_{n \rightarrow \infty} \E[d(X_1^n,\hat{X}_1^n)] \leq D$.
Thus, there exists finite $N$ such a dimension-$N$ quantizer
with $2^{NR}$ codewords achieves distortion at most $D + \epsilon$.
Applying this quantizer to blocks of length $N$ of the source
creates a finite-alphabet source that can be communicated as a
multiset with $O(\log n)$ bits (Theorem~\ref{thm:HRzero}).
The distortion with respect to (\ref{eq:rhodefGeneral}) does not
exceed $D + \epsilon$.
In fact, a somewhat more stringent distortion measure is held to
at most $D + \epsilon$; this measure is of the form (\ref{eq:rhodefGeneral})
with permutations $\pi$ limited to rearrangements that keep blocks of
length $N$ intact.
\end{IEEEproof}

\subsection{Coding of Finite-Size Multisets for Discrete-Alphabet Sources}
\label{sec:finitefinite}
In Sections \ref{sec:entropyrate}, \ref{sec:universal}, and \ref{sec:RDzero}
we allowed the multiset size to go to infinity. Such source coding
incurs infinite delay, and in the case of Section~\ref{sec:RDzero} the
source coding problem is trivialized. 
If we are concerned with delay,
we would want to code short blocks at a time. In this section and
the subsequent section, we investigate bounds on coding when we restrict
the multiset size to be fixed and finite.  Then our asymptotic results
are based on increasing the number of independent realizations of these
finite-sized multisets.

If we are concerned about lossless representation of each fixed multiset,
then the rate requirement is simply lower bounded by the entropy.  
As we had discussed in Section~\ref{sec:entropyrate}, if the multiset elements
are drawn i.i.d., then the entropy is the same as the entropy of a multinomial
random variable.  
For example, Bernoulli$(p)$ multisets of size $K$ have $H(\{X_i\}_{i=1}^K) \approx \tfrac{1}{2} \log_2(2\pi e K p(1 - p))$.
The entropy lower bound assumes that we require the fixed-size multisets to be
uniquely decipherable.  If we insist on a slightly weaker requirement, where these
multisets might become permuted, we can require multiset decipherability of the 
multiset representations.  Notwithstanding the falsity of \cite[Conjecture 3]{Lempel1986} (shown in
\cite{Restivo1989}), the gains below entropy are minimal, and so we do not
pursue this weaker requirement further.

Rather than lossless coding, one might be interested in lossy coding of fixed-size multisets
from discrete alphabets.  We define a fidelity criterion for $K$-size multisets 
\[  
F_{2} = \left\{\frac{K}{n}\sum_{i=1}^{n/K} d_K(x_{iK-K+1}^{iK},y_{iK-K+1}^{iK}), n=K,2K,\ldots\right\}\mbox{.}
\]
The word distortion measure, $d_K$, used to define the fidelity criterion takes value
zero when $x$ and $y$ are in the same type class and one otherwise.  One can also
express the word distortion measure in group theoretic terms using permutation groups, if desired.
This notion of fidelity casts the problem
into a frequency of error framework on the types.
Assuming that the multisets to be coded 
are independent and identically distributed, 
this is 
simply an i.i.d.\ discrete (finite or countable) source with error frequency distortion, so the reverse waterfilling 
solution of Erokhin \cite{Erokhin1958} 
applies.  The rate-distortion 
function is given parametrically as
\begin{align}
D_{\theta} &= 1 - S_{\theta} + \theta(N_{\theta} - 1) \notag \\
R_{\theta} &= -\sum_{\ell:p(\ell)>\theta} p(\ell)\log p(\ell) + (1 - D_{\theta})\log(1 - D_{\theta}) + (N_{\theta} - 1)\theta \log \theta \mbox{,} \notag
\end{align}
where $N_{\theta}$ is the number of types whose probability is greater than $\theta$ and $S_{\theta}$ 
is the sum of the probabilities of these $N_{\theta}$ types.  The parameter $\theta$ goes 
from $0$ to $p(\ell^\ddagger)$ as $D$ goes from $0$ to $D_{\rm max} = 1 - p(\ell^\dagger)$; the most 
probable type is denoted $\ell^\dagger$ and the second most probable type is denoted $\ell^\ddagger$.  
If the letters within the multisets are also i.i.d., the probability values needed
for the reverse waterfilling characterization are computed using the
multinomial distribution.  

Only the most probable source types are used in the representation alphabet.  It is known that the probability 
of type class $k$ drawn i.i.d.\ from the a finite-alphabet parent $p_X$ is bounded as follows \cite{Csiszar1998}:
\[
\tfrac{1}{|\mathcal{K}(\mathcal{X},n)|}2^{-nD(p_k\|p_X)} \le \Pr[\ell] \le 2^{-nD(p_k\|p_X)} \mbox{,}
\]
where $p_k$ is a probability measure derived by normalizing the type $k$.  
The multiset types used in the representation alphabet
are given by the type classes in the typical set
\[
T_{p_X}^{\epsilon(\theta)} = \left\{k: D(p_k\|p_X) \le \epsilon(\theta)\right\} \mbox{.}
\] 
Since multiset 
sources are successively refinable under error frequency distortion \cite{EquitzC1991}, scalable coding 
would involve adding types into the representation alphabet. 

In addition to $F_2$, we can define other fidelity criteria that
reduce the multiset rate-distortion problem to well-known discrete memoryless source rate-distortion
problems.  As a simple example, consider multisets of length $K = 2$ and consisting of i.i.d.\ equiprobable
binary elements.  Then there are three letters in the alphabet of types: 
$\{0,0\}$, $\{0,1\}$, and $\{1,1\}$, which can be represented by their Hamming weights, $\{0,1,2\}$.
The probabilities of these three letters are $\{1/4,1/2,1/4\}$.  Define the word distortion function
using the Hamming weight, $w_H(\cdot)$,:
\[
\delta(x_1^2,y_1^2) = \left| w_H(x_1^2) - w_H(y_1^2)\right| \mbox{.}
\]
The fidelity criterion is
\[
F_{3} = \left\{\frac{2}{n}\sum_{i=1}^{n/2} \delta(x_{2i-1}^{2i},y_{2i-1}^{2i}), n=2,4,\ldots\right\}\mbox{.}
\]
This is a single-letter fidelity criterion on the Hamming weights and is in fact the well-studied problem
known as the Gerrish problem \cite[Problem 2.8]{Berger1971}.  One can easily generate  equivalences to
other known problems as well.

\subsection{Coding of Finite-Size Multisets for Continuous-Alphabet Sources}
\label{sec:finiteuncountable}
Now turning our attention to fixed-size multisets with continuous alphabets,
we first see what simple quantization schemes can do, then develop
some high-rate quantization theory results and finally compute some
rate distortion theory bounds.

\subsubsection{Low-Rate Low-Dimension Quantization for Fixed-Size Multisets}
Using the previously defined word distortion (\ref{eq:rhodef})
on blocks of length $K$,
a new fidelity criterion is 
\[  
F_{4} = \left\{\frac{K}{n}\sum_{i=1}^{n/K} \rho_K(x_{iK-K+1}^{iK},y_{iK-K+1}^{iK}), n=K,2K,\ldots\right\}\mbox{.}
\]
This is average MSE on the block of order statistics.  Notice that the fidelity criterion is 
defined only for words that have lengths that are multiples of the block size $K$.

For low rates and coding one set at a time, we can find 
optimal MSE quantizers through the Lloyd-Max optimization procedure \cite{GershoG1992}.  
The quantizers generated in this way are easy to implement for practical source coding, 
and they also provide an upper bound on the rate-distortion function.
Designing the quantizers requires knowledge of the distributions of order statistics, which can be
derived from the parent distribution \cite{DavidN2003}.  
For $X_1,\,X_2,\,\ldots,\,X_K$ that are drawn i.i.d.\ according to the cumulative
distribution function $F_X(x)$, the marginal cumulative distribution function 
of $X_{(r:K)}$ is given in closed form by 
\[
F_{(r:K)}(x) = \sum_{i=r}^{K}\binom{K}{i}F_X^i(x)\left[1-F_X(x)\right]^{K-i} = I_{F_X(x)}(r,K-r+1)\mbox{,}
\]
where $I_p(a,b)$ is the incomplete beta function.  
Subject to the existence of the parent density $f_X(x)$, the marginal density of
$X_{(r:K)}$ is 
\begin{equation}
\label{eq:margpdf}
f_{(r:K)}(x) = \frac{1}{B(r,K-r+1)}\left[1 - F_X(x)\right]^{K-r} F_X^{r - 1}(x)f_X(x) \mbox{,}
\end{equation}
where $B(a,b)$ is the beta function.  
The joint density of all $K$ order statistics is
\begin{equation}
\label{eq:jointpdf}
f_{(1:K),\ldots,(K:K)}(x_1,\ldots,x_K) =
\left\{ \begin{array}{ll}
  K! \prod_{i=1}^Kf_X(x_i), & x_1^K \in \mathfrak{R}; \\
                        0, & \mbox{else}. \end{array} \right.
\end{equation}
The region of support, $\mathfrak{R} = \{x_1^K: x_1 \le \cdots \le x_K\}$,
is a convex cone that occupies $(1/K!)$th of $\mathbb{R}^K$.
The order statistics also have the Markov property \cite{DavidN2003} with transition probability
\begin{equation}
\label{eq:transpdf}
f_{X_{(r+1:K)}|X_{(r:K)}=x}(y) = (K-r)\left[\frac{1-F_X(y)}{1-F_X(x)}\right]^{K-r-1} \frac{f_X(y)}{1-F_X(x)}, \quad \mbox{for $y>x$.}
\end{equation}

In a standard quantization setup, the sorting filter (\ref{eq:transform})
would be applied first to generate the transform 
coefficients and then further source coding would be performed.  
Since sorting quantized numbers is easier than sorting real-valued numbers,
we would prefer to be able to interchange the operations.
Based on the form of the joint 
distribution of order statistics, (\ref{eq:jointpdf}), we can formulate a statement about 
when sorting and quantization can be interchanged without loss of optimality.  If 
the order statistics are to be quantized individually using scalar quantization, then
interchange without loss can be made in all cases \cite{Gandhi1997}.  Scalar quantization,
however, does not take advantage of the Markovian dependence among elements to be coded.
We consider coding the entire set together, referring to $K$ as the dimension of the order 
statistic vector quantizer.  

\begin{figure}
  \centering
  \includegraphics[width=2.5in]{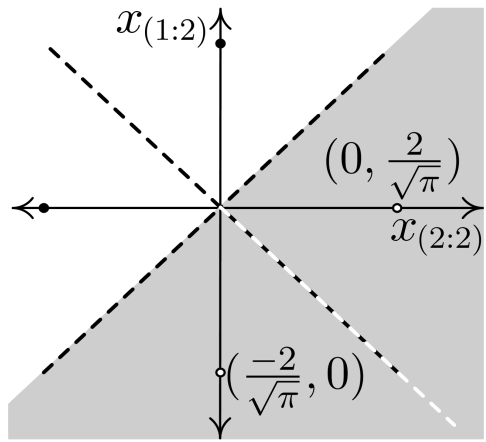}
  \caption[Ordered bivariate Gaussian vector quantizer.]
    {Quantization for bivariate standard Gaussian order statistics.
     Optimal one-bit quantizer (white) achieves $(R = 1,\, D=(2\pi-4)/\pi)$. 
     Optimal two-bit quantizer (black) for unordered variates achieves
     $(R = 2,\, D=(2\pi-4)/\pi)$.  
     Since representation points for order statistic quantizer are the
     intersection of the cone (shaded) and the representation points for
     the unordered quantizer, the distortion performance is the same.}
  \label{fig:gaussOSvq}
\end{figure}

If the representation points for an MSE-optimal 
$(R \mbox{ rate},\, K \mbox{ dimension})$ order statistic 
quantizer are the intersection of $\mathfrak{R}$ with
the representation points for an
MSE-optimal $(R + \log K!,\, K)$ quantizer for the unordered variates,
then we can interchange sorting and quantization
without loss of optimality.  This condition can be interpreted as a
requirement of permutation polyhedral symmetry on the quantizer of
the unordered variates.  This form of symmetry requires that there are corresponding
representation points of the unordered variate quantizer in each of 
the $K!$ convex cones that partition $\mathbb{R}^K$ on the basis of 
permutation.  The polyhedron with vertices that are corresponding points in each
of the $K!$ convex cones is a permutation polyhedron. 
In fact, the distortion performance of the
MSE-optimal $(R,\,K)$ order statistic quantizer is equal to the distortion
performance of the best $(R + \log K!,\, K)$ unordered quantizer constrained 
to have the required permutation symmetry.  An example where the symmetry 
condition is met is for the standard bivariate Gaussian distribution shown in 
Figure~\ref{fig:gaussOSvq}.

\subsubsection{High-Rate Quantization Theory for Fixed-Size Multisets}
\label{sec:highRateQuant}
Based on the basic distributional properties of order statistics, (\ref{eq:margpdf})--(\ref{eq:transpdf}),
the differential entropies of order statistics can be derived.
The individual marginal differential entropies are 
\begin{equation}
h(X_{(r:K)}) = \int f_{(r:K)}(x) \log f_{(r:K)}(x) dx \mbox{,}
\end{equation}
where no particular simplification is possible 
unless the parent distribution is specified.  
The average marginal differential entropy, however, can be expressed in terms of the 
differential entropy of the parent distribution and a constant that depends only on $K$ \cite{WongC1990}:
\begin{equation}
\bar{h}(X_{(1:K)},\ldots,X_{(K:K)}) = \frac{1}{K}\sum_{i = 1}^K h(X_{(i:K)}) = h(X_1) - \log K - \frac{1}{K}\sum_{i = 1}^K \log \binom{K-1}{i-1}  + \frac{K - 1}{2} \mbox{.}
\label{eq:hbar}
\end{equation}
The subtractive constant is positive and increasing in $K$, and not dependent on the parent distribution.  
The individual conditional differential entropies, as derived in \cite{Varshney2006}, are
\begin{align}
&h(X_{(r+1:K)}|X_{(r:K)}) = -\log(K-r) - N_h(K) + N_h(K-r) +1 - \frac{1}{K-r} \\ \notag
&- \frac{K!}{\Gamma(K-r)\Gamma(r)} \int_{-\infty}^{\infty} \int_{x}^{\infty} f_X(y)\log(y) \left[1 - F_X(y)\right]^{K-r-1} dy F_X^{r-1}(x) f_X(x) dx \mbox{,}
\end{align}
where $\Gamma(\cdot)$ is the gamma function and $N_h(k) = \sum_{m=1}^k 1/m$ is the harmonic number.  
As in the individual marginal case, further simplification of this expression requires
the parent distribution to be specified.  
Again, as in the marginal case, the total conditional
differential entropy can be expressed in terms of the parent differential entropy and
a constant that depends only on $K$.  Due to Markovianity, the sum of the individual conditional 
differential entropies is in fact the joint differential entropy:
\begin{equation}
h\left(X_{(1:K)},\ldots,X_{(K:K)}\right) = h(X_{(1:K)}) + \sum_{i=1}^{K-1} h(X_{(i+1:K)}|X_{(i:K)}) = Kh(X_1) - \log K! \mbox{.}
\end{equation}
Notice that an analogous statement (\ref{eq:orderlowerbound}) was a lower bound in the 
discrete alphabet case; equality holds in the continuous case since there are no ties.  

High-rate quantization results follow easily from the differential entropy calculations.
To develop results, we introduce four quantization 
schemes in turn, measuring performance under fidelity criterion $F_{4}$.  
In particular, we sequentially introduce a \emph{shape advantage},
a \emph{memory advantage}, and a \emph{space-filling advantage} as in~\cite{LookabaughG1989}.\footnote{Note 
that vector quantizer advantages are discussed in terms of 
distortion for fixed rate in \cite{LookabaughG1989}, but we present some of these advantages 
in terms of rate for fixed distortion.}
As a baseline, take the na\"{i}ve scheme of direct uniform scalar quantization
of the arbitrarily-ordered sequence with quantization step size $\epsilon$. 
The average rate and distortion 
per source symbol of the na\"{i}ve scheme are $R_1 = h(X_1) - \log \epsilon$, and $D_1 = \epsilon^2/12$.  
Now instead uniformly scalar quantize the deterministically-ordered
sequence (the order statistics).  This changes the shape of the marginal distributions 
that we are quantizing, and thus we get a 
shape advantage. 
The average rate per source symbol for this scheme is
\begin{equation}
R_2 = \bar{h}(X_{(1:K)},\ldots,X_{(K:K)}) - \log \epsilon = R_1 - \log K - \frac{1}{K}\sum_{i = 1}^K \log \binom{K-1}{i-1}  + \frac{K - 1}{2}\mbox{.}
\end{equation}
The distortion is the same as the na\"{i}ve scheme, $D_2 = D_1$. 
As a third scheme, 
scalar quantize the order statistics sequentially, using the previous 
order statistic as a form of side information.  Even though the encoding 
requires waiting for the entire block so as to sort, decoding can proceed
symbol-by-symbol, reducing delay.
We assume that the previous order statistics 
are known exactly to both the encoder and decoder.  
Since the order statistics form 
a Markov chain, this single-letter sequential transmission exploits all available memory advantage.  
The rate for this scheme is
\begin{equation}
R_3 = \tfrac{1}{K}h(X_{(1:K)},\ldots,X_{(K:K)}) - \log \epsilon = R_1 - \tfrac{1}{K}\log K!\mbox{.}
\end{equation}
Again, $D_3 = D_1$.  Finally, the fourth scheme would vector quantize the entire sequence of 
order statistics collectively.  Since we have exploited all shape and memory advantages, 
the only thing we can gain is 
space-filling gain.  The rate is the same as the third
scheme, $R_4 = R_3$, however the distortion is less.  This distortion reduction
is a function of $K$, is related to the best packing of polytopes, and 
is not known in closed form for most values of $K$;
see~\cite[Table~I]{LookabaughG1989} 
and more recent work on packings. 
We denote the distortion as $D_4 = D_1/G(K)$, where $G(K)$ is a 
function greater than unity.  The performance improvements of these schemes 
are summarized in Table~\ref{tab:advantage}.  Notice that all values in Table~\ref{tab:advantage} 
depend only on the multiset length $K$ and not on the parent distribution.
\begin{table}
  \centering
    \begin{tabular}{|l|c|c|}
      \hline
      & Rate Reduction $(-)$ & Distortion Reduction $(\times)$ \\ \hline
      Scheme 1 & $0$ & $1$ \\ \hline
      Scheme 2 (s) & $\log K + \frac{1}{K}\sum_{i = 1}^K \log \binom{K-1}{i-1}  - \frac{K - 1}{2}$ & $1$ \\ \hline
      Scheme 3 (s,m) & $(\log K!)/K$ & $1$ \\ \hline
      Scheme 4 (s,m,f) & $(\log K!)/K$ & $1/G(K)$ \\ \hline
    \end{tabular}
  \caption[Rate and distortion performance for several high-rate quantization schemes.]{Comparison between the na\"{i}ve scalar quantization (Scheme 1)
and several 
other quantization schemes.  The symbols (s), (m), and (f) denote shape, memory, and space-filling 
advantages.}
  \label{tab:advantage}
\end{table}

We have introduced several quantization schemes and calculated their performance in the high-rate limit.
It was seen that taking the fidelity criterion into account when designing the source coder resulted
in rate savings that did not depend on the parent distribution.  
These rate savings can be quite significant for large blocklengths $K$.  

\subsubsection{Rate Distortion for Fixed-Size Multisets}
It is quite difficult to obtain the full rate-distortion function 
for the $F_{4}$ fidelity criterion;
however, upper and lower bounds may be quite close to each other for
particular source distributions.  As an example, consider the 
rate-distortion function for the independent bivariate standard Gaussian 
distribution that was considered in Figure~\ref{fig:gaussOSvq}.  The rate-distortion
function under $F_{4}$ is equivalent to the rate-distortion 
function for the order statistics under the MSE fidelity criterion, as shown.  
For clarity of expression, let $X = (X_i)_{i=1}^K$ and $\hat{X} = (Y_i)_{i=1}^K$ in the 
unsorted domain and $Z = \{X_i\}_{i=1}^K$ and $\hat{Z} = \{Y_i\}_{i=1}^K$ in the 
sorted domain.  Clearly, the fidelity constraint is naturally expressed in the 
$Z$ domain.  The affirmatively answered question is whether the mutual information 
in the rate distortion optimization can be switched from $I(X;\hat{Z})$ to $I(Z;\hat{Z})$:
\begin{align}
I(X;\hat{Z}) &= h(X) + h(\hat{Z}) - h(X,\hat{Z}) \\ \notag
		 &\stackrel{(a)}{=} h(Z) + h(J) + h(\hat{Z}) - h(X,\hat{Z}) \\ \notag
		 &= h(Z) + h(J) + h(\hat{Z}) - \left[h(X,\hat{Z}|J) + I(X,\hat{Z};J)\right] \\ \notag
		 &= h(Z) + h(J) + h(\hat{Z}) - h(Z,\hat{Z}) - I(X,\hat{Z};J) \\ \notag
		 &= h(Z) + h(J) + h(\hat{Z}) - h(Z,\hat{Z}) - h(J) + h(J|X,\hat{Z}) \\ \notag
		 &\stackrel{(b)}{=} I(Z;\hat{Z}) + h(J) - h(J) = I(Z;\hat{Z}) \mbox{.}
\end{align}
Step (a) is due to Theorem~\ref{thm:Hdecomp} and step (b) follows since 
$h(J|X,\hat{Z})$ is zero.

The Shannon lower bound is simply 
\begin{equation}
R_{\rm SLB}(D) = \log(1/D),
\end{equation}
the Gaussian rate-distortion function under the MSE fidelity criterion, reduced 
by $\log K!$ bits (one bit).  Note that since the order statistic source cannot be written as the 
sum of two independent processes, one of which has the properties of a Gaussian with variance 
$D$,\footnote{Even though $X_{(1:2)} = \tfrac{1}{2}(X_1 + X_2) - \tfrac{1}{2}|X_1 - X_2|$ and 
$X_{(2:2)} = \tfrac{1}{2}(X_1 + X_2) + \tfrac{1}{2}|X_1 - X_2|$, 
and the first terms are Gaussian, the troublesome part is the independence.} the Shannon lower bound
is loose everywhere \cite{GerrishS1964}, though it becomes asymptotically tight in the high-rate limit.  

\begin{figure}
  \centering
  \includegraphics[width=3.0in]{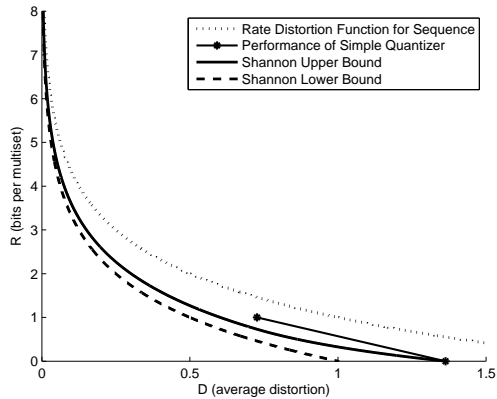}
  \caption{Shannon upper and lower bounds for the 
Gaussian order statistic rate-distortion function.  The point achievable by single set code of 
Figure~\ref{fig:gaussOSvq} is also shown connected to the zero rate point, which is known to be tight.  Note 
that rate is not normalized per source letter.}
  \label{fig:rdorder2}
\end{figure}

The covariance matrix of the Gaussian order statistics can be computed
in closed form as 
\begin{equation}
\Lambda = \left[ {\begin{array}{*{20}c}
   {1 - 1/\pi} & {1/\pi}  \\
   {1/\pi} & {1 - 1/\pi}  \\
\end{array}} \right] \mbox{,}
\end{equation}
with eigenvalues $1$ and $1 - 2/\pi$. 
Reverse waterfilling yields the Shannon upper bound
\begin{equation}
R_{\rm SUB}(D) = \left\{ {\begin{array}{*{20}c}
   {\tfrac{1}{2}\log\left(\frac{2 - 4/\pi}{D}\right) + \tfrac{1}{2}\log\left(\frac{2}{D}\right),} & {0 \le D \le 2 - 4/\pi }  \\
   {\tfrac{1}{2}\log\left(\frac{1}{D - 1 + 2/\pi}\right),} & {2 - 4/\pi \le D \le 2 - 2/\pi}  \\
   {0,} & {D \ge 2 - 2/\pi.}  \\
\end{array}} \right.
\end{equation}
This bound is tight at the point achieved by zero rate.  Since the Gaussian order
statistics for $K=2$ have small non-Gaussianity, the Shannon lower bound and the Shannon upper bound
are close to each other, as shown in Figure~\ref{fig:rdorder2}.  For moderately small distortion values, we
can estimate the rate-distortion function quite well.

The fact that the Shannon lower bound is loose everywhere applies not only to
the particular example we considered, but to any problem.  That is to say, $\log(K!)$
bits cannot be saved below the rate distortion function for the usual squared error fidelity
criterion. 
\begin{thm}
\label{thm:SLBloose}
The Shannon lower bound to the rate distortion function is loose everywhere 
for any source, under the fidelity criterion $F_4$.
\end{thm}
\begin{IEEEproof}
The support of the joint distribution (\ref{eq:jointpdf}) 
for any order statistic source is the convex cone $\mathfrak{R} = \{x_1^K: x_1 \le \cdots \le x_K\}$.
The support of a Gaussian distribution is all of $\mathbb{R}^K$.
For the Shannon lower bound to be tight, the source must be decomposable as the 
sum of two independent processes, one of which has the properties of a Gaussian 
\cite{GerrishS1964}.  Since the Gaussian density has support over all space, it cannot be 
convolved with another density (non-negative) to yield a 
third density that has support over only part of space.
\end{IEEEproof}

\section{Universal Lossy Coding}
\label{sec:universalLossy}
The final setting in which we investigate the ramifications of
order irrelevance is universal lossy coding.
The general goal in universal source coding is to find encoding algorithms
that perform well for all members of a class of sources~\cite{NeuhoffGD1975}.
Here we have the modest goal of demonstrating that
$O(\log n)$ rate requirements extend quite generally to lossy coding.
The results we present are not intended to be conclusive,
but rather are included to wind up our tour of source coding.

Recall the main result of Section~\ref{sec:RDzero}:
under a \emph{per-letter} MSE fidelity criterion,
zero distortion is achievable with zero \emph{total} rate for a large
class of sources.  This result is obtained with the number of letters $n$
growing without bound and the source distribution known.
An interpretation of this is that using the known distribution to
pseudorandomly ``simulate'' the source at the destination is sufficient
for achieving zero distortion.
Now we consider a universal setting in which this approach will not
work because the source distribution is not known at the destination.

Instead of giving a result for real-valued sources and multiset MSE,
we jump directly to a more general result.
Let $d$ be a single-letter distortion measure, and
let $R^*(D)$ denote an operational rate distortion function
that is achievable by fixed-rate codes uniformly over a class of sources
(coding the sources as sequences).
As in Section~\ref{sec:lossyArbitrary},
for coding without regard to order,
consider the single-letter distortion measure $\rho_n$
and associated fidelity criterion $F_1$
given in (\ref{eq:rhodefGeneral})--(\ref{eq:F1defB}).
Denote by $R^*_{ \{X_i\}_{i=1}^n }(D)$
the minimum (total) rate for encoding $\{X_i\}_{i=1}^n$
with $\E[\rho_n(X_1^n,\hat{X}_1^n)] \leq D]$ for every source in the class.
Then we obtain the following result analogous to Theorem~\ref{thm:lossy_logn}:
\begin{thm}
\label{thm:lossyUniversal}
If $R^*(D)$ is finite, then for any $\epsilon > 0$, 
\[
R^*_{\{X_i\}_{i=1}^n}(D + \epsilon) = O(\log n) \mbox{.}
\] 
\end{thm}
\begin{IEEEproof}
Let $D$ be such that $R = R^*(D)$ is finite and let $\epsilon > 0$.
The achievability of $R^*(D)$ \emph{uniformly over the class}
means that there is a finite dimension $N$ at which distortion
$D+\epsilon$ is achieved at rate $R$ for every source in the class.
Thus only minor adjustments to the proof of Theorem~\ref{thm:lossy_logn}
are needed.
\end{IEEEproof}

As a simple application consider a set of real-valued parent distributions
that share a bounded support.
An arbitrarily small multiset MSE can be obtained uniformly over all
the sources with $O(\log n)$ rate.
This follows from Theorem~\ref{thm:lossyUniversal} because the finiteness
of $R^*(D)$ for any positive distortion $D$ can be demonstrated by
uniform quantization of the support of the source class.

\section{Concluding Comments: From Sequences to Multisets}
\label{sec:seq2set}
We have completed a tour through the major areas of source coding while
discussing how things are changed by irrelevance of the order of source letters.
To conclude, we discuss three conceptual transitions between sequences
and multisets and then summarize.

\subsection{Types $\rightarrow$ Markov Types $\rightarrow$ Sequences}
\label{sec:Markovtypes}
In the Shannon-style language approximations that were mentioned in the 
opening, a first-order approximation corresponds to a multiset of letters
of the original alphabet, whereas an approximation of the same order
as the length of the sequence is the sequence itself. 
In between these extremes, there are many possibilities:
A second-order approximation is a multiset of digrams
(ordered pairs of source letters),
a third-order approximation is a multiset of trigrams, etc.
Thus, as the approximation order is increased, the lengths 
of segments within which the ordering of letters is relevant increases.

\begin{figure}
\centering
\includegraphics[width=3.0in]{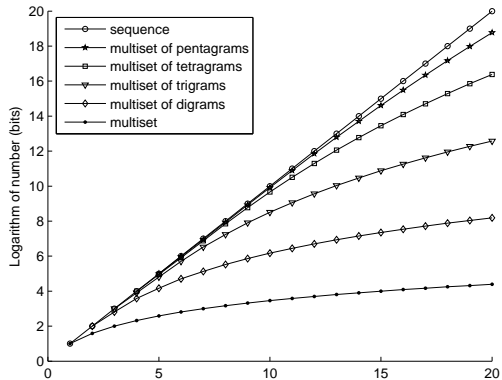}
\caption{Logarithm of the number of binary sequences, multisets, or $k$-gram multisets as a function of the number of binary letters.}
\label{fig:Sets_rate}
\end{figure}

For a fixed alphabet, increasing the approximation order also causes
the number of distinct outcomes to increase. 
For first-order approximations to $n$ source letters drawn from
alphabet $\mathcal{X}$,
the number of distinct outcomes is
the number of types.
For an $\ell$th-order approximation,
the number of distinct outcomes is the number of \emph{Markov type
classes}~\cite{DavissonLS1981,Shields1990,JacquetS2004}. 
Markov type classes are also known as
\emph{kinds} in cognitive science~\cite{GriffithsT2001,GriffithsT2007}
and used to visualize motifs in computational genomics~\cite{SchneiderS1990}. 
The enumeration of Markov types is not simply expressed \cite{JacquetS2004}, but
can be upper bounded by $(n + 1)^{|\mathcal{X}|^{\ell}}$.
The number of Markov 
types gives an upper bound on the rate requirements for lossless coding and
is computed exactly in Figure~\ref{fig:Sets_rate}.  
There is no source that can achieve the enumeration upper bound for different values 
of $\ell$ simultaneously since it is impossible to have equiprobable 
sequences and multisets at the same time.  For a real source, 
like the empirical source from the Zenith radio experiments 
in telepathy~\cite{Goodfellow1938}, the entropy is much lower than the 
bounds; see Table \ref{table:zenith}.

\begin{table}
\caption{Entropy of Zenith Radio Telepathy Data}
\label{table:zenith}
\centering
\begin{tabular}{|c|c|c|}
\hline
& \small{Entropy (bits)} & \small{Bound (bits)}\\
\hline
\small{sequence} & \small{4.6663}& \small{5.0000}\\
\hline
\small{multiset of tetragrams} & \small{4.5892}& \small{4.9542}\\
\hline
\small{multiset of trigrams} & \small{4.3359}& \small{4.8074}\\
\hline
\small{multiset of digrams} & \small{3.5411}& \small{4.1699}\\
\hline
\small{multiset} & \small{1.8758}& \small{2.5850}\\
\hline 
\end{tabular}
\end{table}
               
\subsection{Partially Commutative Alphabets}
Rather than varying ordering requirements by varying the segments
over which order is relevant, one can allow particular letters in the alphabet 
to commute in position with other letters.  
As discussed in~\cite{Savari2004}, a source with such a partially-commutative
alphabet can be described by a noncommutation graph.  As edges are removed
from this graph, the importance of the order of letters decreases.
In the case of the empty noncommutation graph, the order is irrelevant
and the so-called lexicographic normal form associated with the 
noncommutation graph is simply the sequence sorted 
into order.  The distinct outcomes associated with a noncommutation graph
are called interchange classes and the moment-generating function 
for the number of interchange classes is equal to the inverse of 
the M\"{o}bius polynomial corresponding to a function of the 
noncommutation graph.  Interchange entropies are discussed in detail
by Savari~\cite{Savari2004}.

If noncommutation graphs are defined on sliding windows of source letters
rather than on individual letters, the problem becomes
one of source coding with a $0$-$1$ context-dependent fidelity 
criterion \cite{Shannon1959,BergerY1972}; sliding windows that have 
distortion zero between them commute.  Since the sliding windows overlap,
however, the commutation relations must be constrained to remain consistent.
Just as sources with partially-commutative alphabets
lead to type classes when the noncommutation graph is empty, sources with empty
noncommutation graphs on sliding windows lead to Markov type classes.  

\subsection{Quantum Physics}
\label{sec:quantum}
In statistical physics, the Maxwell-Boltzmann statistics are used for 
non-interacting, identical bosons in the classical limit and correspond
to sequences, whereas the Bose-Einstein statistics are used 
when quantum effects are manifested and correspond to
multisets.  In the classical regime, bosons of the same energy 
level, $x \in \mathcal{X}$, may be distinguished by their different 
positions in space.  That is to say the order of particles is important.
As the concentration of particles increases, some particles become
so close that they can no longer be distinguished in
position, and degeneracy results.  Thus
degeneracy measures the importance of order in 
representing bosons.  When the concentration of particles
exceeds the quantum concentration, i.e.\ when the 
interparticle distance is less than the thermal de Broglie wavelength,
the bosons become indistinguishable and so representable by a multiset.
The combinatorics of indistinguishability as a function of particle 
concentration is given in the so-called partition function for bosons.
The partition function allows distributional characterizations of intermediate 
levels of particle concentration to be made.

Incidentally, at even greater concentrations than the quantum concentration,
the probability mass of letters appearing in the multiset concentrates 
on a single letter, as determined by the average boson occupation number. 
This is known as Bose-Einstein condensation.  

\subsection{Summary}
Partial or full order irrelevance has significant qualitative impact.
For lossless coding of $n$ letters from a finite-alphabet source,
the rate requirement grows only logarithmically with $n$
rather than linearly with $n$ (Theorem~\ref{thm:HRzero});
the rate reduction as a ratio is thus arbitrarily large.
In a universal setting, the rate reduction is again arbitrarily large:
for a source satisfying
Kieffer's condition for sequence representation~\cite{Kieffer1978},  
universal coding with $n + o(n)$ bits is achievable
(Theorem~\ref{thm:universalAchievability}).
This should be compared with $cn$ bits when order is relevant,
where constant $c$ could be arbitrarily large.
Despite this positive statement about universal coding,
it is impossible for the redundancy to be a negligible fraction
of the coding rate (Section~\ref{sec:universalMemorylessMultisets}).

For lossy coding subject to per-letter MSE distortion,
irrelevance of order can trivialize the source coding problem
for a large class of sources.  Specifically, under rather weak moment
conditions on the parent distribution, zero distortion is achieved
even with zero rate as $n \rightarrow \infty$
(Theorem~\ref{thm:zerozero}).
This is not of practical importance because a source coder will
process only a finite amount of data at once.
High-resolution analyses of various quantization schemes for a
block of size $K$ are presented in Section~\ref{sec:highRateQuant}.
Through the inclusion of shape and memory advantage---and under
the assumption of high rate---a rate savings of $\log K!$ bits can
be achieved relative to na\"{i}ve scalar quantization.
However, in a rate distortion setting the ``full'' savings of
$\log K!$ bits can only be achieved as the rate approaches infinity,
not at any finite rate (Theorem~\ref{thm:SLBloose}).

\section*{Acknowledgments}
The authors thank Alon Orlitsky for fruitful discussions;
in particular, the results in Section~\ref{sec:achievability} were developed
in collaboration with him. 
The authors also thank Sanjoy K. Mitter for several discussions.

\bibliographystyle{IEEEtran} 
\bibliography{abrv,lrv_lib}

\end{document}